# **A**utomatic **R**ecommendation **f**or **O**nline **U**sers **U**sing **W**eb **U**sage **M**ining

Ms.Dipa Dixit [1] Mr Jayant Gadge[2]

Lecturer [1] Asst.Professor[2]

Fr CRIT , Vashi Navi Mumbai[1]

Thadomal Shahani Engineering College,Bandra [2]

Email:dipa.pathak@gmail.com [1]

jayantrg@hotmail.com[2]

## *ABSTRACT*

*A real world challenging task of the web master of an organization is to match the needs of user and keep their attention in their web site. So, only option is to capture the intuition of the user and provide them with the recommendation list. Most specifically, an online navigation behavior grows with each passing day, thus extracting information intelligently from it is a difficult issue. Web master should use web usage mining method to capture intuition. A WUM is designed to operate on web server logs which contain user's navigation. Hence, recommendation system using WUM can be used to forecast the navigation pattern of user and recommend those to user in a form of recommendation list. In this paper, we propose a two tier architecture for capturing users intuition in the form of recommendation list containing pages visited by user and pages visited by other user's having similar usage profile. The practical implementation of proposed architecture and algorithm shows that accuracy of user intuition capturing is improved.*

## 1. INTRODUCTION

With the explosive growth of knowledge available on World Wide Web, which lacks an integrated structure or schema, it becomes much more difficult for users to access relevant information efficiently. Meanwhile, the substantial increase in the number of websites presents a challenging task for web masters to organize the contents of websites to cater to the need of user's. Analyzing and modeling web navigation behavior is helpful in understanding demands of online users. Following that, the analyzed results can be seen as knowledge to be used in intelligent online applications, refining website maps, and web based personalization system and improving searching accuracy when seeking information. Nevertheless, an online navigation behavior grows each passing day, thus extracting information intelligently from it is a difficult issue. Web Usage Mining (WUM) is process of extracting knowledge from Web user's access data, by exploiting Data Mining technologies. It can be used for different purposes such as personalization, system improvement and site modification. A typical application of Web Usage Mining is represented by so called recommender system. The main goal of the recommender system is to improve Web site usability. Typically, the Web usage mining prediction process is structured according to two components performed online and off-line with respect to Web server activity. Offline component builds the knowledge base by analyzing historical data, such as server access log file or web logs which are captured from the server, then these web logs are used in the online component for capturing the intuition list of the user so as to recommend page views to the user whenever he / she comes online for the next time.

In our paper, we present architecture for capturing recommendations in the form of intuition list for user. Intuition List consists of list of pages visited by user as well as list of pages visited by other user of having similar usage profile**.** The results represent that improved accuracy of recommendations. The rest of this paper is organized as follows: In section 2, we review some researches that advance in understanding of recommendation systems using web usage mining. Section 3 describes the block diagram and implementation for the Recommendation System. Results and discussion are shown in section 4. Finally, section 5 summarizes the paper and introduces future work.

## 2. RELATED WORK

Recently, several Web Usage Mining systems have been proposed to predicting user navigation behavior and their preferences. In the following we review some of the most significant WUM systems and architecture that can be compared with our system Analog[8] is one of the first WUM systems .It is structured according to an offline and an online component. The off-line component build session clusters by analyzing past user activity recorded in server log files. Then the online component builds active user sessions which are then classified according to generated model. The classification allows to identify pages related to the ones in the active session and to return the requested page with a list of suggestions. This approach has several limitations, related to scalability. Nevertheless, architectural solution introduced was maintained in several other more projects. In Mobasher et al[1] present Web personalizer a system which provides dynamic recommendations, as a list of hypertext links, to users. The analysis is based on anonymous usage data combined with the structure formed by hyperlinks of the site. Data mining techniques (i.e. clustering, sequence pattern discovery and association rules) are used in preprocessing phase in order to obtain aggregate usage profiles. In this phase Web server logs are converted into clusters of visited pages, and cluster made up of set of pages with common usage characteristics. The online phase considers active user session in order to find matches among user's activities and discovered usage profiles. Matching entries are used to compute a set of recommendations which will be inserted into last requested page as list of hypertext links. Web Personalizer is a good example of two tier architecture for Personalization Systems. Baraglia and Palmerini proposed a WUM system called SUGGEST, that provide useful information to make easier the web user navigation and to optimize the web server performance [6, 7]. SUGGEST adopts a two level architecture composed of offline creation of historical knowledge and online engine that understands user's behavior. As the request arrives at this system module it incrementally updates a graph representation of web site based on the active user sessions and classifies the active session using a graph partitioning algorithm. Potential limitation of this architecture might be:
a) the memory required to store Web server pages in quadratic in the number of pages .This might be severe limitation in larger sites made of million pages;
b)it does not permit us to manage web sites made up of pages dynamically generated. All of these works attempt to find the architecture and algorithm to improve accuracy of personalized recommendation, but accuracy still does not meet satisfaction. In our work we advance architecture and propose a classification approach using visited and unvisited pages of user in the architecture for improving accuracy of recommendation for users.

# 3. BLOCK DIAGRAM AND IMPLEMENTATION OF RECOMMENDATION SYSTEM

Block diagram of the Recommendation System is given below.

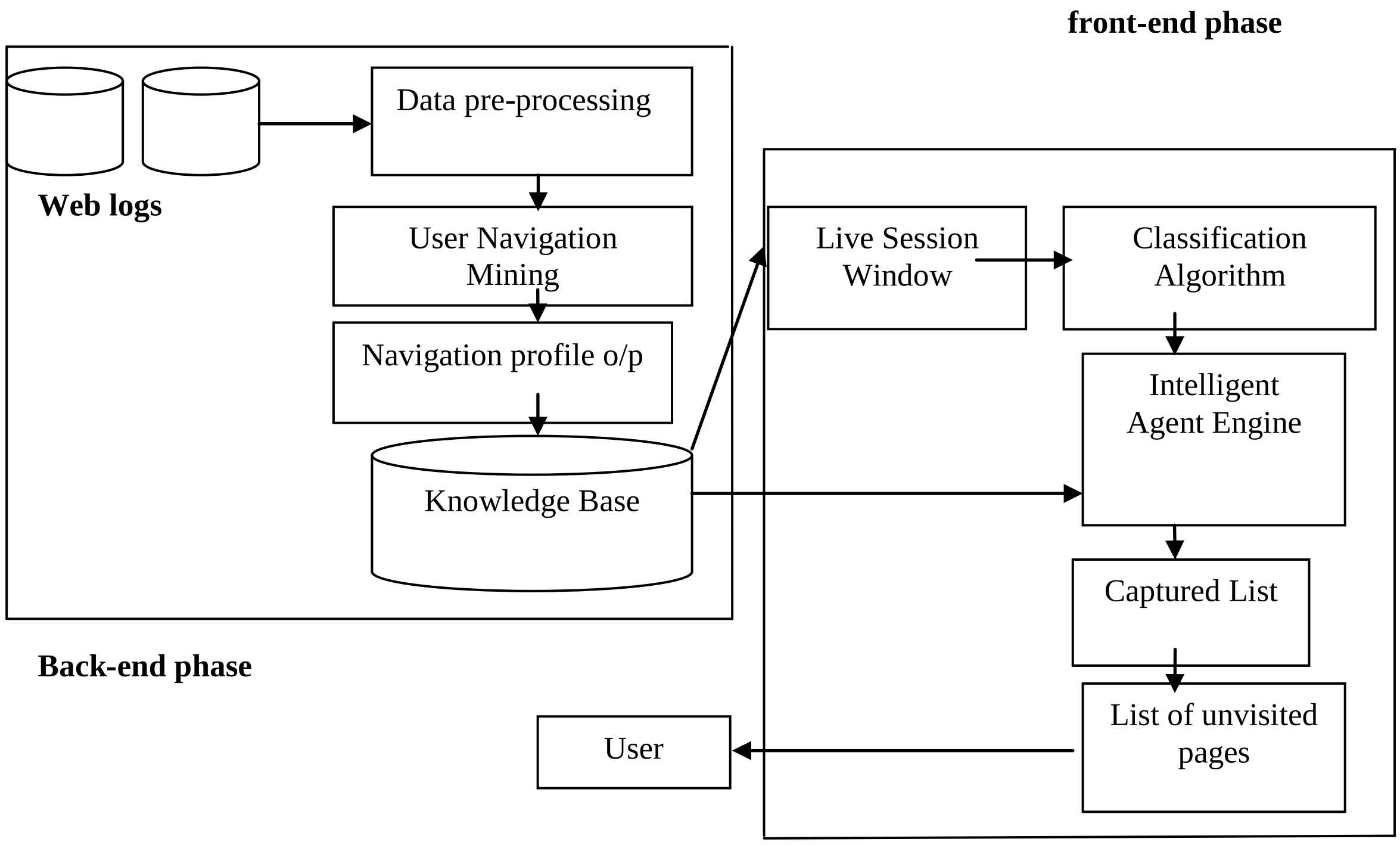

Figure1: Architecture of Recommendation System for Online Users

## 3.1 Implementation of Recommendation System

Implementation of System is done in two phases, Back-end and Front-end phase.

**3.1.1Back-end phase :**

Steps involved in back-end phase are explained below.

**Step 1:** Data sets consisting of 5000 web log records are collected from De Paul University website. Web log is an unprocessed text file which is recorded from the IIS Web Server. Web log consist of 17 attributes with the data values in the form of records.

Fragment of web log from IIS web server is shown below:

Fields: date time c-ip cs-username s-sitename s-computername s-ip s-port cs-method cs-uri-stem cs-uri-query sc-status time-taken cs-version cs-host cs(User-Agent) cs(Referer) .

**Step2:** Generally, several preprocessing tasks need to be done before performing web mining algorithms on the Web server logs. Data preprocessing, a web usage mining model aims to reformat the original web logs to identify user's access session. The Web server usually registers all users' access activities of the website as Web server log. Due to different server setting parameters, there are many types of web logs, but typically the log files share the same basic information, such as: client IP address, request time, requested URL, HTTP status code, referrer, etc.

Data pre-processing is done using following steps.

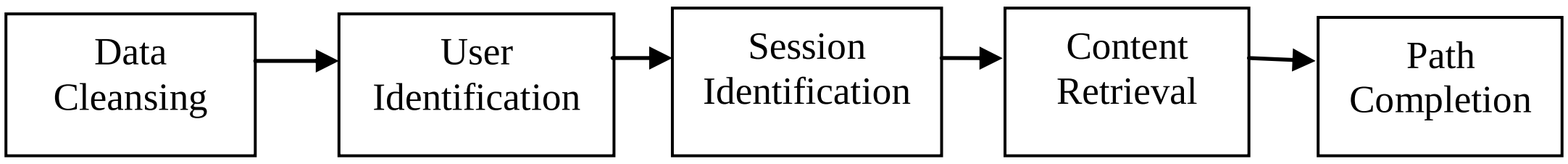

Figure 2: Block diagram for Pre-processing

1. **Data Cleansing**: Irrelevant records are eliminated during data cleansing. Since target of web usage mining is to get traversal pattern, following two kinds of records are unnecessary and should be removed :

   a. The records having filenames suffixes of GIF, JPEG, CSS.

   b. By examining the status field of every record in the web log, the record with status code over 299 and below 200 are removed**.**

2. **User and Session Identification**: The task of user and session identification is to find out the different user sessions from the original web access log. A referrer-based method is used for identifying sessions. The different IP addresses distinguish different users.

   a. If the IP addresses are same, different browsers and operation system's indicate different users which can be found by client IP address and user agent who gives information of user's browsers and operating system.

   b. If all of the IP address, browsers and operating systems are same, the referrer information should be taken into account. The ReferURI is checked, new user's session is identified if the URL in the ReferURI is '-' that is field hasn't been accessed previously, or there is a large interval of more than 30 minutes between the accessing time of this record.

3. **Content Retrieval:** Content Retrieval retrieves content from users query request i.e. cs_uri_query.
Eg:Query:http/1www.cs.depaul.edu/courses/syllabus.asp?course=323-21-603&q=3&y=2002&id=671.

   Retrieve the content like /courses/syllabus.asp which helps in fast searching of unvisited pages i.e; pages of other user's which are similar to user's interest.

4. **Path Completion:** Path Completion should be used acquiring the complete user access path. The incomplete access path of every user session is recognized based on user session identification. If in a start of user session, Referrer as well URI has data value, delete value of Referrer by adding '-'. Web log preprocessing helps in removal of unwanted records from the log file and also reduces the size of original file by 40-50%.

**Step 3: Generation of Page Id:** Page Id is sequence generated numbers like p1, p2, p3….which are created for pages/page views.

**Step 4: User Navigation Mining:** Web pages accessed are modeled as undirected graph G= (V, E). The set V of vertices contains the identifiers of the different pages hosted on the Web server and E is edges of the graph.

a. Undirected graph is created for a single user session using Hash Map.

b. Hash Map data structure stores the referrer–URI pair and their corresponding weights.

c. Weight of edges given by 1, only if link between page and referrer exist, else weight is 0.

d. Weights of pages are frequency connectivity of pages in graph. i.e.

**Weight of pages (W) = Frequency (F) of referrer–URI pair (**occurrence in user session**)**

e. Apply Depth First Search Algorithm (DFS) on graph and obtain all possible navigation patterns.

f. Path length of pattern is calculated by considering the total weight of the edges in a graph.

g. If navigation pattern weight /path length is less than three, then pattern is not considered for analysis. (Minpathlength = 3).

Hence, clusters of patterns for user sessions are obtained and fed into Knowledge base for further analysis.

### 3.1.2 Front End Phase

**Step 1: Longest Common Subsequence Algorithm:**

a. Capture the Live Session Window (LSW) for a user dynamically [5].

b. Intelligent Agent Engine compares LSW of a user with patterns of same user in knowledge base.

c. Check for the longest pattern or the largest path length of a pattern from knowledge base.

d. Compare both the sequences, longest common subsequence is obtained.

e. Consider the pages which are not present in subsequence, these pages are the Intuition pages for the user as they are visited by user most frequently.

f. Recommendation list is given in the form of URI (content) as well as IDs of pages.

Hence, recommendation/intuition list as compared to user's historical pattern are captured.

**Step2: Searching of Unvisited pages (**as compared to others user's pattern**)**

a. Unvisited pages when compared to other user's pattern are searched using searching algorithm.

b. Searching algorithm compares the live session window of user and patterns of other user's present in Knowledge Base.

c. Best possible pattern is achieved by considering longest path length (weight).

d. Subsequent pages are removed and the Unvisited Page List is created in the form of URI and IDs.

Recommendation List as compared to others user's pattern (unvisited pages) are captured and added to original recommendation list

## 4. RESULTS AND DISCUSSIONS

Step wise results are shown below for 5000 web log records from De Paul University dataset (CTI dataset).

**Step 1:** Collection of web logs which are in raw or unprocessed form.17 attributes are shown below:

2002-04-01 00:00:10 1cust62.tnt40.chi5.da.uu.net - w3svc3 bach bach.cs.depaul.edu 80 get /courses/syllabus.asp course=323-21-603&q=3&y=2002&id=671 200 156 http/1.1 www.cs.depaul.edu mozilla/4.0+(compatible;+msie+5.5;+windows+98;+win+9x+4.90;+msn+6.1;+msnbm sft;+msnmen-us;+msnc21) http://www.cs.depaul.edu/courses/syllabilist.asp depaul.edu/courses/syllabilist.asp

2002-04-01 00:00:26 ac9781e5.ipt.aol.com - w3svc3 bach bach.cs.depaul.edu 80 get /advising/default.asp - 200 16 http/1.1 www.cs.depaul.edu mozilla/4.0+(compatible;+msie+5.0;+msnia;+windows+98;+digext) http://www.cs.depaul.edu/news/news.asp?theid=573

**Step 2:** Preprocessing is done for 5000 web log records. Cleansing, User and Session Identification, Content Retrieval and Path Completion applied on records.

Preprocessing of 5000 records was done in **14secs.**

User Sessions were identified for 200 users.

Thus, processed records for a user id 9 in user's sessionized form are as shown below.

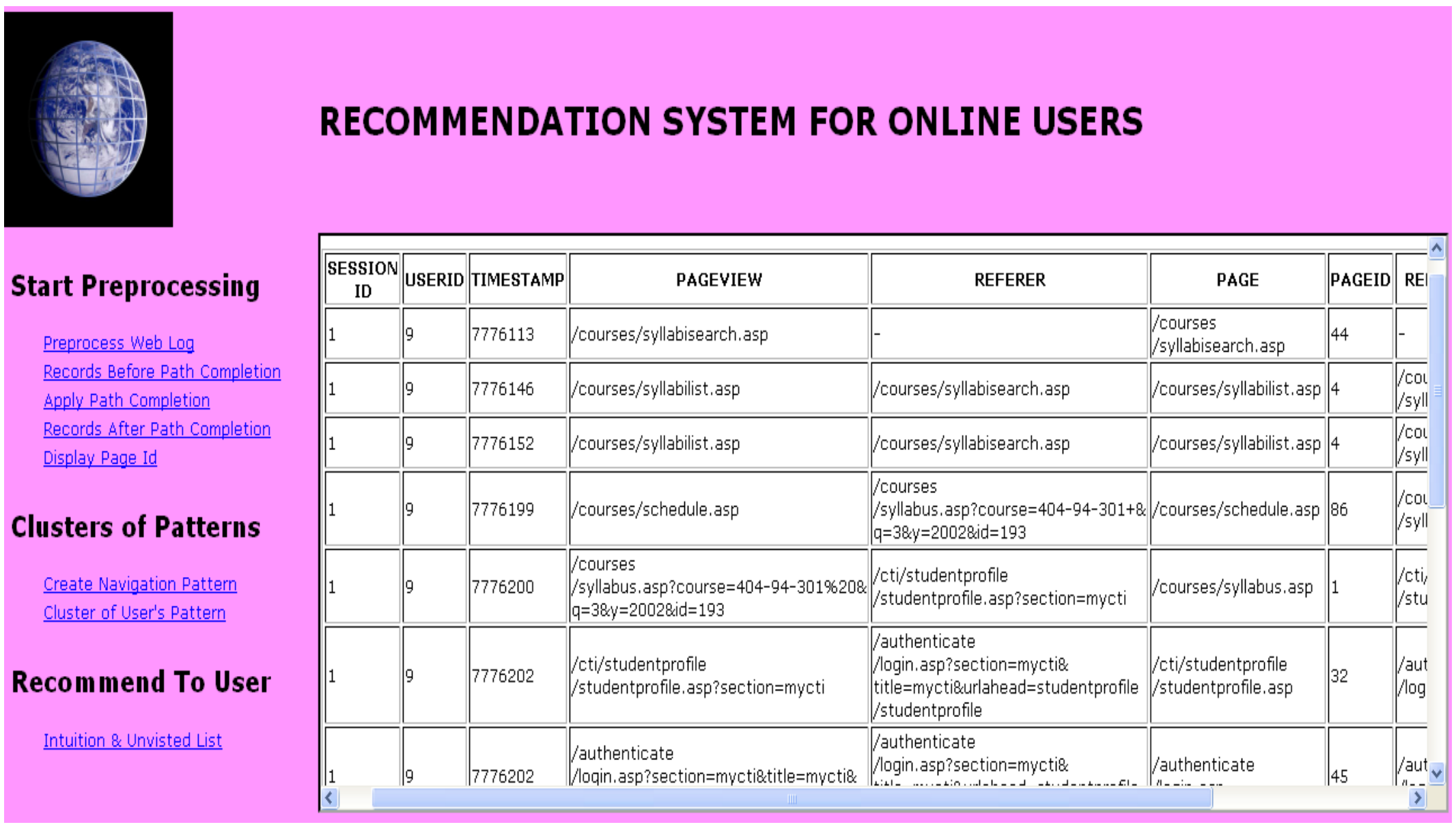

Figure 3: Processed file with required attributes for user id 9

**Step 3:** Page Id is generated for the URI/pages/page view accessed by user

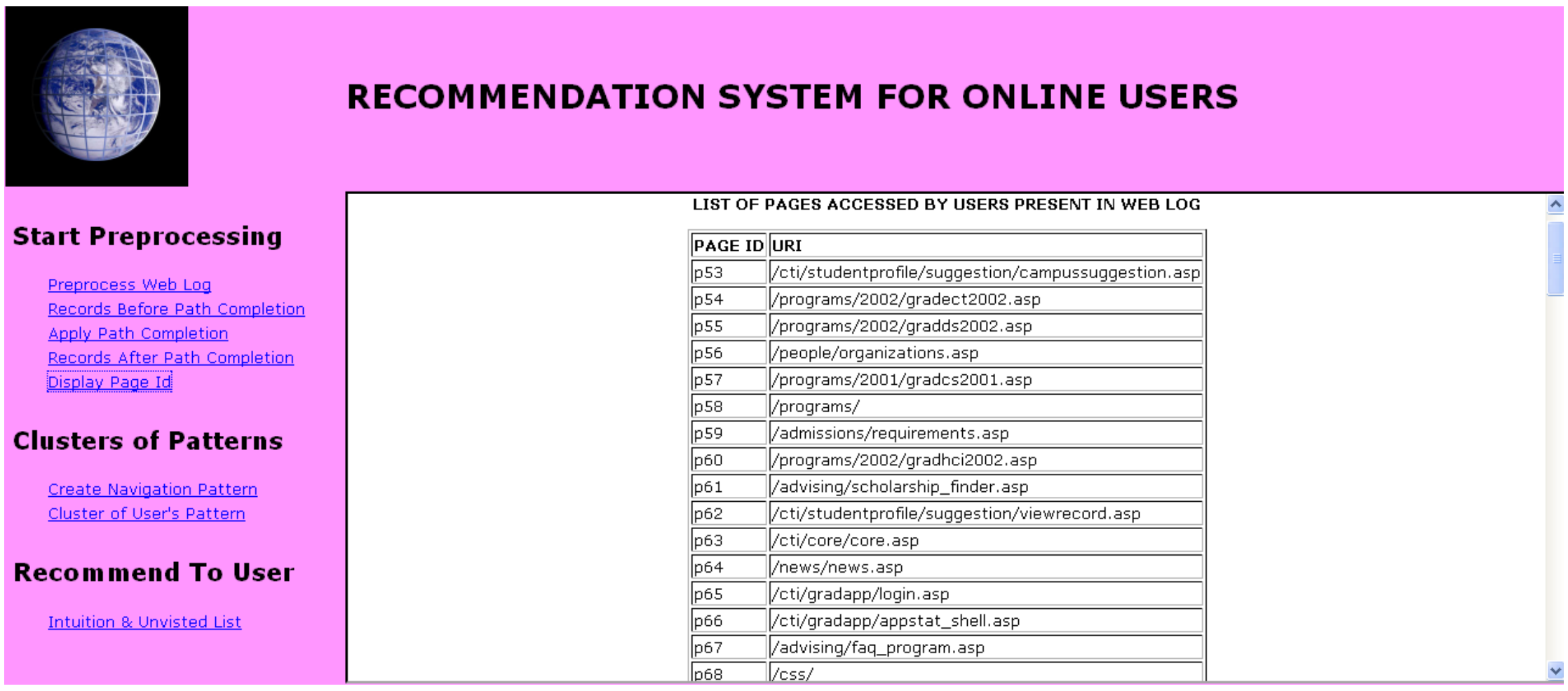

| PAGE ID | URI |
|---|---|
| p53 | /cti/studentprofile/suggestion/campussuggestion.asp |
| p54 | /programs/2002/gradect2002.asp |
| p55 | /programs/2002/gradds2002.asp |
| p56 | /people/organizations.asp |
| p57 | /programs/2001/gradcs2001.asp |
| p58 | /programs/ |
| p59 | /admissions/requirements.asp |
| p60 | /programs/2002/gradhci2002.asp |
| p61 | /advising/scholarship_finder.asp |
| p62 | /cti/studentprofile/suggestion/viewrecord.asp |
| p63 | /cti/core/core.asp |
| p64 | /news/news.asp |
| p65 | /cti/gradapp/login.asp |
| p66 | /cti/gradapp/appstat_shell.asp |
| p67 | /advising/faq_program.asp |
| p68 | /css/ |

Figure 4: List of Page id and corresponding pages/uri

**Step 4:** In User Navigation Mining undirected graphs are created and clusters of all possible patterns are generated for a user session

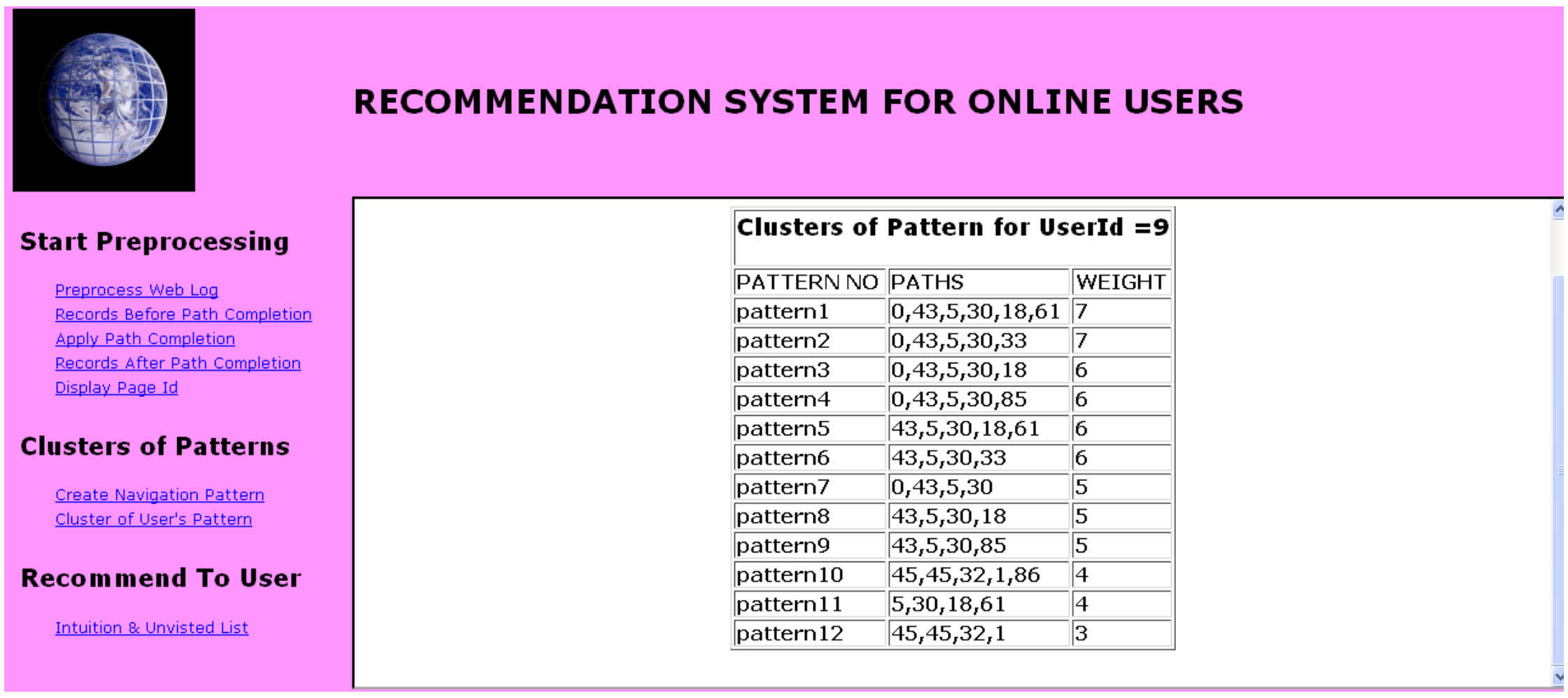

| PATTERN NO | PATHS | WEIGHT |
|---|---|---|
| pattern1 | 0,43,5,30,18,61 | 7 |
| pattern2 | 0,43,5,30,33 | 7 |
| pattern3 | 0,43,5,30,18 | 6 |
| pattern4 | 0,43,5,30,85 | 6 |
| pattern5 | 43,5,30,18,61 | 6 |
| pattern6 | 43,5,30,33 | 6 |
| pattern7 | 0,43,5,30 | 5 |
| pattern8 | 43,5,30,18 | 5 |
| pattern9 | 43,5,30,85 | 5 |
| pattern10 | 45,45,32,1,86 | 4 |
| pattern11 | 5,30,18,61 | 4 |
| pattern12 | 45,45,32,1 | 3 |

Figure 5: Cluster of patterns for user id=9

When Clusters of navigation Patterns were compared to original user navigation patterns most of pages views (99%) were covered in the clusters. Very few i.e; 1% of outliers were obtained.

**Step 5:** Considered Two cases of Live Session Window (LSW) of size 2 and varying patterns ie; one pattern having few page views and other having more number of page views these cases are shown in Step 6. Live Session Window (LSW) consists of 10% of pages of actual page views of user ($Original_{np}$). Classification is done by applying Longest Common Subsequence algorithm on LSW and rest of pages present in original page view list of user. Thus the intuition list obtained is from the history of user's navigation pattern.

**Step 6:** Apply Searching algorithm to get the intuition list of the user's whose usage pattern is same as the user. Thus, both the lists are combined into the Final Recommendation List of the user. Finally, Accuracy is calculated for the final recommendation list.
Accuracy measures the degree to which the recommendation system produces accurate recommendations. It is given by

$$\frac{|P(LS_{np}, lsw) \cap (Original_{np})|}{|P(LS_{np}, lsw)|} \quad (1)$$

lsw = Live Session Window

$P(LS_{np}, lsw)$ -Navigation pattern in captured list recommended by engine.

$Original_{np}$ -Original page views / pattern of user.

**Case 1:** When LSW of size 2 was considered for user id 9, having 13 page views. Recommendation List obtained had accuracy of 66.6%, which is shown below.

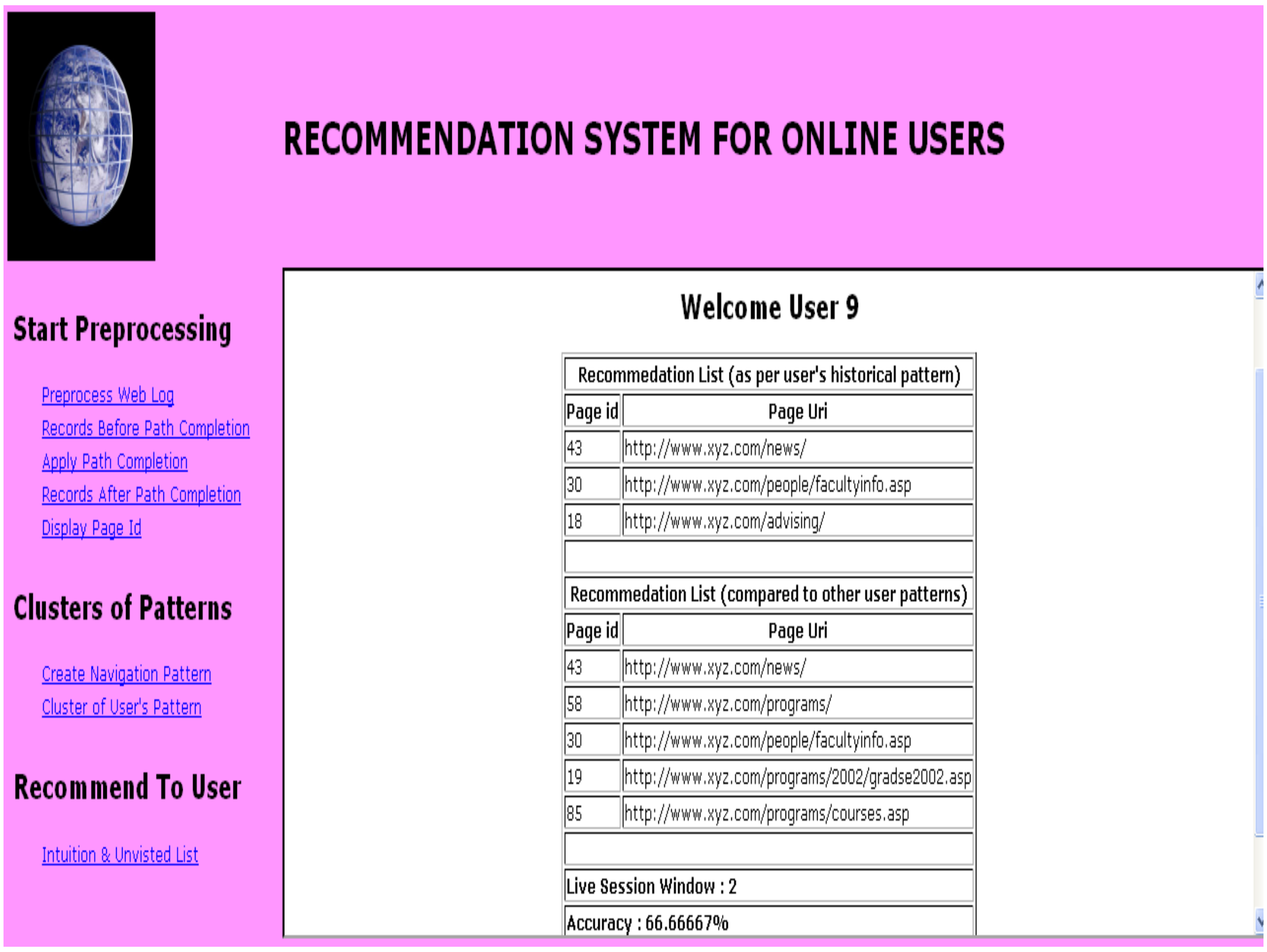

Figure 6: recommendation list for user id=9 and accuracy is 66.6%

**Case 2:** When LSW of size 2 was considered for user id 89, having 17 page views, recommendation list obtained had accuracy 85.71%, as shown below.

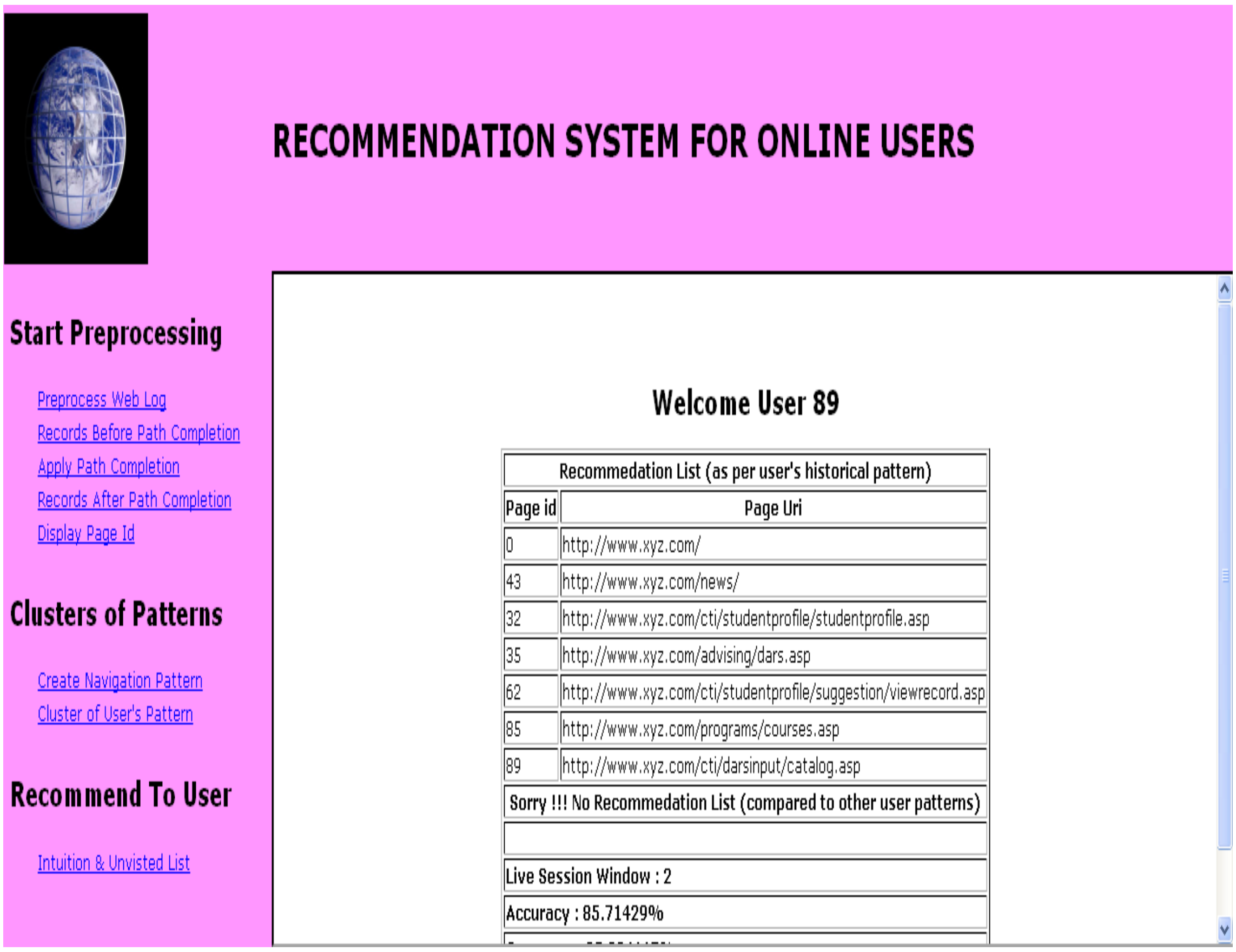

Figure 7: recommendation list for user id=89 and accuracy is 85.71%

Thus, from above cases we can prove that accuracy of the recommendation list increases if the number of page views is more in the user navigation pattern.

## 5. CONCLUSION

In this paper, we propose a two tier architecture for capturing user's intuition in the form of recommendation list containing list of pages visited by user and also list of pages visited by other user's having similar usage profile. The practical implementation of proposed architecture and algorithms shows that accuracy of user intuition capturing improves up to 85 percent for Live Session Window size of two, if numbers of page views having maximum weights are more in the navigation patterns of the user. In the future, we would like to substantially improve the accuracy and coverage parameter by trying to increase the Live Session Window (LSW) size and considering more number of log records.